\documentclass[a4paper,11pt]{article}
\pdfoutput=1 

\usepackage{jheppub} 

\usepackage[T1]{fontenc} 

\title{\boldmath Analysis of black hole thermodynamics under Generalized uncertainty principle from the doubly special relativity}

\author[a]{E. Maghsoodi,}
\author[a,1]{H. Hassanabadi,\note{Corresponding author.}}
\author[b]{and Won Sang Chung}

\affiliation[a]{Faculty of Physics, Shahrood University of Technology, Shahrood, Iran P. O. Box: 3619995161-316.}
\affiliation[b]{Department  of  Physics and Research Institute of Natural Science,
	College of Natural Science,
	Gyeongsang National University, Jinju 660-701, Korea}

\emailAdd{e.maghsoodi184@gmail.com}
\emailAdd{h.hasanabadi@shahroodut.ac.ir}
\emailAdd{mimip44@naver.com}

\abstract{In this paper, we investigate effect of the generalized uncertainty principle (GUP) under the doubly special relativity (DSR) on the thermodynamics properties of the topological charged black hole in Anti-de Sitter (AdS) space only in the spherical horizon case have. Our study is based on a heuristic analysis of a particle which is captured by the black hole. We also report here some results of a usual analytical computation. However, we obtain the black hole thermodynamic properties as familiar concepts such as temperature, entropy and heat capacity under DSR-GUP. we also compare our both analytical results. Beside we discuss the behavior of the corrected thermodynamic properties vs. changes of black hole characteristics under different condition.}

\begin{document} 
\maketitle
\flushbottom

\section{Introduction}
\label{sec:intro}
In quantization of gravity, the concept of minimum measurable length is included when considering the discreteness of the space time which occurs beyond the Planck energy scale refs.~\cite{1,2,3,4,5,6}. By considering $\hbar  = 1$ , the ordinary Heisenberg uncertainty principle is given by $\Delta x\Delta p \geqslant \frac{1}{2}$  that it does not include minimum measurable length because in the high momentum limit, $\Delta x$ goes to zero. 

Therefore, deforming the ordinary Heisenberg uncertainty is a clear way to import the concept of minimum measurable length into quantum mechanics. Thus one can improve the ordinary Heisenberg uncertainty principle which is called Generalized uncertainty principle (GUP). This principle will lead to the corrected the thermodynamic quantities refs.~\cite{71,72,018,73,74,8,9,10,11,12,13,14,15,16,18,19,20,21,22,23,24,25}. These quantities have been successfully considered to describe and investigate behavior of black holes. 

Recently, by embracing two invariant fundamental scales, the speed of light and an energy scale distinguished with planck energy, the deformation of the special relativity was considered ref.~\cite{6}. Because this theory has two invariant scale, this modified form of theory is usually called a doubly special relativity (DSR).

In the last decade, the new form of GUP based on DSR was introduced and in this manuscript we propose a new form of GUP as follows
\begin{eqnarray}
\left[ {\hat x,\hat p} \right] = i{\left( {1 - \frac{{\left| {\hat p} \right|}}{\kappa }} \right)^2} ,\label{s00}
\end{eqnarray} 
This gives
\begin{eqnarray}
\Delta \hat x\Delta \hat p \geqslant \frac{1}{2}\left[ {1 - 2\frac{{\left\langle {\left| {\hat p} \right|} \right\rangle }}{\kappa } + \frac{1}{{{\kappa ^2}}}{{\left( {\Delta \hat p} \right)}^2}} \right]
\label{s01}
\end{eqnarray}
Where from now on we call the relation \eqref{s01} DSR-GUP. The expectation value of   in Eq. (\ref{s01}) deepens on the wave function and this value runs from $0$  to $\dfrac{\kappa}{2}$ ref.~\cite{6}. Thus \eqref{s01} can be rewritten as
\begin{eqnarray}
\Delta \hat x\Delta \hat p \geqslant \frac{1}{2}\left[ {1 - 2a + \frac{1}{{{\kappa ^2}}}{{\left( {\Delta \hat p} \right)}^2}} \right],
\label{s02}
\end{eqnarray}
Where $0 < a < \frac{1}{2}$. The different between \eqref{s00} and the defined form of GUP  as   $\left[ {\hat x,\hat p} \right] = i\left( {1 + \beta {{\hat p}^2}} \right)$ is the momentum operator representation. Recently in usual form, the effects of quantum gravity into fermions’ tunneling from Reissner-Nordstrom and Kerr black holes ref.~\cite{73} and the effects of the minimal length on the motion of a particle perturbed away from an unstable equilibrium near the black hole horizon ref.~\cite{74} have studied. 

In this paper we consider the deformed derivative corresponding to DSR-GUP to investigate the quantum mechanics based on algebraically way and besides using usual analytical computation we have considered a very useful heuristic method for a particle which is captured by the black hole. In this method, a black hole appears as a sink for capturing particles that the black hole's size and temperature are obtained via $\Delta x$ and $\Delta p$ parameters and this method is based on that the black hole mass is not allowed to be less than a scale order plank mass.

The obtained solutions for the thermodynamic properties depend on the black hole's characteristics such as black hole's mass, charge, pressure, the cosmological constant and et al. and it depends on considered general topology. 

This topology is also depending on the cosmological constant $\Lambda $  and it has different asymptotic behavior,  this can be asymptotically flat  $\left( {\Lambda  = 0} \right)$, de Sitter $\left( {\Lambda  > 0} \right)$ or anti-de Sitter $\left( {\Lambda  < 0} \right)$. In this paper, we consider the charged anti-de Sitter (AdS) black hole with special topology "spherical horizon".

This paper is organized as follows: In section~\ref{sec:A class of static and spherically black holes} we discuss the topology of black holes under DSR-GUP. In section~\ref{sec:Black hole thermodynamics} we discuss the mass-temperature relation for black hole and in section~\ref{sec:Black hole thermodynamics under DSR-GUP} we have calculated the DSR-GUP corrected thermodynamic quantities of the charged black hole and discuss their properties and some numerical analyses. Finally, a conclusion is presented in section~\ref{sec:Conclusions}.

\section{A class of static and spherically black holes}
\label{sec:A class of static and spherically black holes}
In the present paper, we would like to investigate the topological black holes in the four-dimensional space-time.  We consider a static and spherical black hole black hole solution as follows  refs.~\cite{018,73,74}
\begin{eqnarray}
d{S^2} =  - F(r)d{t^2} + {F^{ - 1}}(r)d{r^2} + {r^2}d{\Omega ^2}, \label{s3}
\end{eqnarray}
where $r$  is cosmological radius and $d{\Omega ^2}$ is the line element of a two-dimensional hypersurface $\Omega$ with constant curvature $2K$ 
\begin{eqnarray}
d{\Omega ^2} =\begin{cases}
d{\theta ^2} + {\sin ^2}\theta \,d{\phi ^2} ~~~~~~~~ \text{for}~ K = 1
\\     d{\theta ^2} + {\theta ^2}\,d{\phi ^2} ~~~~~~~~~~~\text{for}~ K = 0
\\     d{\theta ^2} + {\sinh ^2}\theta \,d{\phi ^2} ~~~~~~\text{for}~ K = -1
\end{cases}
\end{eqnarray}
and the metric function $F(r)$ define as
\begin{eqnarray}
F(r) = K - \frac{{8\pi GM}}{{{\Sigma _K}r}} + \frac{{16{\pi ^2}{G^2}{Q^2}}}{{\Sigma _K^2{r^2}}} + \frac{{{r^2}}}{{{\ell ^2}}},\label{s4}
\end{eqnarray}
where  $M$,  ${\Sigma _K}$  and  $Q$  have identified mass, volume and electric charge and $\ell$  is cosmological radius and $\Lambda=-3\ell^{-2}$ is the negative cosmological constant.
Here for simplicity, we have considered $\frac{{4\pi G}}{{{\Sigma _K}}} = 1$ .
Without loss of generality,  we can consider $K = 1$  for spherical horizon, $K = 0$  for planar/toroidal horizon and $K =  - 1$  for  hyperbolic horizon.

According to the metric function in \eqref{s4}   with considering $K = 1$ and  $F({r_0}) = 0$ , the black hole mass is
\begin{eqnarray}
\label{g1}
M = \frac{{{r_0}}}{2} + \frac{{{Q^2}}}{{2{r_0}}} - \frac{\Lambda }{6}r_0^3,
\end{eqnarray}
Where ${r_0}$  denotes the position of the event horizon of the black hole. By defining  the thermodynamic pressure as $P =  - \Lambda /8\pi $, Then the surface gravity of the black hole is obtained as follows ref.~\cite{18}
\begin{eqnarray}
\kappa  = \frac{{F'({r_0})}}{2} = \frac{{ - \,{Q^2} + r_0^2 + 8\pi Pr_0^4}}{{2r_0^3}},
\end{eqnarray}
In this research, we study the spherical horizon case  and  
in this case, the solution \eqref{s3} is the Reissner-Nordstrom-anti-de Sitter black hole spacetime and the
event horizon has the topology $S^{2}$.

\section{Black hole thermodynamics}
\label{sec:Black hole thermodynamics}
\hspace{0.5cm}
From the first law of black hole mechanics and classical general relativity, we can define 
\begin{eqnarray}
dM = \frac{\kappa }{{8\pi }}dA + \Phi dQ + VdP,\label{s1}
\end{eqnarray}
where $V$  stands for the thermodynamic volume of the black hole and $P$  is the thermodynamic pressure, and  $M$,   $T$,  $S$, $\Phi$ and $Q$  are respectively the black hole mass, temperature, entropy, electric potential, and charge. 

By introdused topology in before section, we can identify the thermodynamic volume as $V=\dfrac{4}{3}\pi r_{0}^{3} $  and the obtained electric potential  at infinity with reference to the horizon as $\Phi  = \frac{Q}{{{r_0}}}$.  However, from the thermodynamic point of view and with respect to the Hawking, the above equation can be rewritten as follows 
\begin{eqnarray}
dM = TdS + \Phi dQ + VdP,
\end{eqnarray}
Generally, black hole entropy should be a function of the horizon area ref.~\cite{26}. Following from \eqref{s1} and according to thermodynamic principles , the temperature of a black hole can be generally calculated as 
\begin{eqnarray}
T = {\left( {\frac{{\partial M}}{{\partial S}}} \right)_Q} = \frac{{dA}}{{dS}} \times {\left( {\frac{{\partial M}}{{\partial A}}} \right)_Q} = \frac{\kappa }{{8\pi }} \times \frac{{dA}}{{dS}},
\end{eqnarray}
In the semi-classical case, the temperature and entropy for the black hole are
\begin{eqnarray}
{S_0} = \frac{A}{{4\hbar }},\,\,\,\,\,\,\,\,\,\,\,\,{T_0} = \frac{{\kappa \hbar }}{{2\pi }}.\label{e1}
\end{eqnarray}
When the particle absorbs by black hole, the smallest increase in the area of a black hole can be considered as ref.~\cite{25}
\begin{eqnarray}
A \sim Xm, \label{s2}
\end{eqnarray}
by defining $X$  and $m$  as the particle$^{,}$s size and mass, respectively and By considering ${\left( {\Delta S} \right)_{\min }} = \rm Ln 2$. ref.~\cite{25} and by Knowing that in quantum mechanics,
the momentum uncertainty is not allowed to be greater than the mass (${\Delta p \leqslant m}$), 
we have studied a particle by a wave packet that the width of wave packet is decribed as the
standard deviation of $X$ distribution i.e. the position uncertainty, which can be
identified as the characteristic size of the particle (${X \sim \Delta x}$). 
Thus  the representation \eqref{s2} can be rewritten as follows
\begin{eqnarray}
A \sim Xm \geqslant \Delta x\Delta p.\label{s6}
\end{eqnarray}
When a particle is captured by black hole, $\Delta x$ should not be greater than a specific scale which minimizes $\Delta A$. This characteristic size should be related to the black hole, if $\Delta A_{\rm min}$ is expected to describe an intrinsic property of the horizon. 

\section{Black hole thermodynamics under DSR-GUP}
\label{sec:Black hole thermodynamics under DSR-GUP}
For considered topology as a static and spherically symmetric black hole, it's size is identified with the twice radius of horizon, i.e.
\begin{eqnarray}
2{r_0} \geqslant \Delta x \geqslant \frac{1}{2}\left[ {\frac{{1 - 2a}}{{\Delta p}} + \frac{{\Delta p}}{{{\kappa ^2}}}} \right],\,\label{s4}
\end{eqnarray}
Therefore From \eqref{s02}, the momentum uncertainty is obtained as
\begin{eqnarray}
2r_0^{}{\kappa ^2} - \kappa \sqrt {4r_0^2{\kappa ^2} + 2a - 1}  \leqslant \Delta p \leqslant 2r_0^{}{\kappa ^2} + \kappa \sqrt {4r_0^2{\kappa ^2} + 2a - 1} ,\label{s5}
\end{eqnarray}
Substituting \eqref{s4} and \eqref{s5} into \eqref{s6}, the increase in area satisfies
\begin{eqnarray}
A \geqslant \gamma \hbar ',\label{s7}
\end{eqnarray}
where $\gamma $ is a calibration factor and we have
\begin{eqnarray}
\hbar ' = 4r_0^2{\kappa ^2} - 2{r_0}\kappa \sqrt {4r_0^2{\kappa ^2} + 2a - 1} ,\label{s8}
\end{eqnarray}
here $\hbar '$  is the effective Planck constant and is considered as  \eqref{s8}.
Thus we can see, DSR-GUP changes the semiclassical framework to a certain context, and the semiclassical black hole temperature \eqref{e1} should satisfy a the DSR-GUP corrected black hole temperature 
\begin{eqnarray}
\label{g2}
T' &=& \frac{{\kappa \hbar '}}{{2\pi }} = \frac{1}{{4\pi r_0^7}}{\left( {{Q^2} - 8\pi Pr_0^4 - r_0^2} \right)^2}  \nonumber\\
&\times &\left[ {8\pi Pr_0^4 - {Q^2} + r_0^2 - \sqrt {64{\pi ^2}{P^2}r_0^8 + 16\pi Pr_0^6 + 2ar_0^4 - 16\pi P{Q^2}r_0^4 - 2{Q^2}r_0^2 + {Q^4}} } \right],
\end{eqnarray}
Since DSR-GUP only constrains the minimal length, while the electric charge and the electric potential will remain unchanged, it correction only influences the temperature and the entropy. The first law of black hole thermodynamics $dM = T'dS' + \Phi dQ + VdP$  should still be established in this case. Therefore from  usual analytical computation, the DSR-GUP corrected entropy of the black hole can be derived as
\begin{eqnarray}
\label{g3}
S' = \int {{{\left. {\frac{{dM}}{T'}} \right|}_{Q,P}}}  = \int {\frac{1}{T'}} {\left. {\frac{{\partial M}}{\partial r }} \right|_{Q,P}}dr ,
\end{eqnarray}
In other hand, according to Heisenberg uncertainty principle, one can derive
\begin{eqnarray}
\frac{{dA}}{{dS}} = \frac{{{{\left( {\Delta A} \right)}_{\min }}}}{{{{\left( {\Delta S} \right)}_{\min }}}} = 4\hbar ',
\end{eqnarray}
Where $\hbar '$  is the effective Planck “constant” and is defined as \eqref{s8}. By defining $\gamma  = 4\operatorname{Ln} 2$, from introdused the heuristic analysis, the DSR-GUP corrected entropy of the black hole can be calculated
\begin{align}
\label{g4}
S' &= \int {\frac{{dS'}}{{dA}}} dA \simeq \int {\frac{{{{\left( {\Delta S'} \right)}_{\min }}}}{{{{\left( {\Delta A} \right)}_{\min }}}}} dA \simeq \int {\frac{{dA}}{{4\hbar '}}}   \nonumber \\ 
&= \frac{{ - 1}}{{8P\left( {2a - 1} \right)}}\left( {8\pi Pr_0^2 + \sqrt {64{\pi ^2}{P^2}r_0^4 + 16\pi Pr_0^2 + 2a} } \right)  \nonumber \\ 
&+ \frac{1}{{8P{{\left( {2a - 1} \right)}^{1/2}}}}\left( {Ln\left( {16\pi } \right) + Ln\left( {\frac{{\sqrt {2a - 1} \sqrt {64{\pi ^2}{P^2}r_0^4 + 16 \pi Pr_0^2 + 2a}  + 2aP - P}}{{8\pi Pr_0^2 + 1}}} \right)} \right)
\end{align}
By considering \eqref{g1} and \eqref{g2} in \eqref{g3}, black hole entropy is obtained as \eqref{g4}.  Therefore with two methods, we have received the same result for uncharged black hole entropy.
Here the effect of DSR-GUP leads to a logarithmic term, which also exists in many other quantum corrected entropy. As a thermodynamic system, the thermodynamic quantities of the black hole should satisfy the thermodynamic identity.

In the semi-classical case, the heat capacity for black hole is given by
\begin{eqnarray}
{C} = {T}\frac{{\partial {S}}}{{\partial {T}}} = \kappa {\left( {4\hbar '} \right)^{ - 1}}\frac{{\partial A}}{{\partial \kappa }},
\end{eqnarray}
Now the heat capacity of the DSR-GUP black hole from usual analytical computation can be deflned as
\begin{eqnarray}
C' = T'\frac{{\partial S'}}{{\partial T'}} = T'\frac{{{\raise0.7ex\hbox{${\partial S'}$} \!\mathord{\left/
				{\vphantom {{\partial S'} {\partial {r_0}}}}\right.\kern-\nulldelimiterspace}
			\!\lower0.7ex\hbox{${\partial {r_0}}$}}}}{{{\raise0.7ex\hbox{${\partial T'}$} \!\mathord{\left/
				{\vphantom {{\partial T'} {\partial {r_0}}}}\right.\kern-\nulldelimiterspace}
			\!\lower0.7ex\hbox{${\partial {r_0}}$}}}},
\end{eqnarray}
and from introduced the heuristic analysis, the DSR-GUP corrected heat capacity can be found as
\begin{align}
	C' = T'\frac{{\partial S'}}{{\partial T'}} = \frac{{\kappa \hbar '}}{{2\pi }} \cdot \frac{{\partial S'}}{{\partial A}} \cdot \frac{{\partial A}}{{\partial T'}} = \frac{{\kappa \hbar '}}{{2\pi }} \cdot \frac{{{{\left( {\Delta S'} \right)}_{\min }}}}{{{{\left( {\Delta A} \right)}_{\min }}}} \cdot \frac{{\partial A}}{{\partial T'}},
\end{align}
	we also can write as follows
	\begin{eqnarray}
	C' = \frac{\kappa }{4}\left( {\frac{1}{\kappa }\frac{{\partial A}}{{\partial \hbar '}} + \frac{1}{{\hbar '}}\frac{{\partial A}}{{\partial \kappa }}} \right), 
	\end{eqnarray}
	By introducing  $\lambda  = \sqrt {{{\left( {8\pi Pr_0^4 + r_0^2 - {Q^2}} \right)}^2} + 2ar_0^4 - r_0^4} $  in the context of the DSR-GUP, calculation for the heat capacity gives
	\begin{align}
	C' &= \frac{1}{4}{\left( {\frac{{\partial \hbar '}}{{\partial A}} + \frac{{\hbar '}}{\kappa }\frac{{\partial \kappa }}{{\partial A}}} \right)^{ - 1}}\,  \nonumber  \\
	&= \left( {2\pi \lambda r_0^6} \right)\left[ { - 7{Q^6} + 3{Q^4}r_0^2\left( {5 + 24\pi Pr_0^2} \right) - 2{Q^2}r_0^4\left( {2 + 5a + 24\pi Pr_0^2\left( {1 - 4\pi Pr_0^2} \right)} \right)} \right. \nonumber \\   
	&  + 2r_0^6\left( {a - 24a\pi Pr_0^2 - 32{\pi ^2}{P^2}r_0^4\left( {9 + 40\pi Pr_0^2} \right)} \right) - 7{Q^4}\lambda  + 8{Q^2}\lambda r_0^2 - r_0^4\lambda  + 16\pi P{Q^2}\lambda r_0^4 \nonumber  \\
	& {\left. { + 32\pi P\lambda r_0^6 + 320{\pi ^2}{P^2}\lambda r_0^8} \right]^{ - 1}}. 
	\end{align}
	One of the main results is that, we have received the same result for charged black hole heat capacity from two methods.
	
\begin{figure}[tbp]
	\centering 
	\includegraphics[height=7cm,width=9cm,clip]{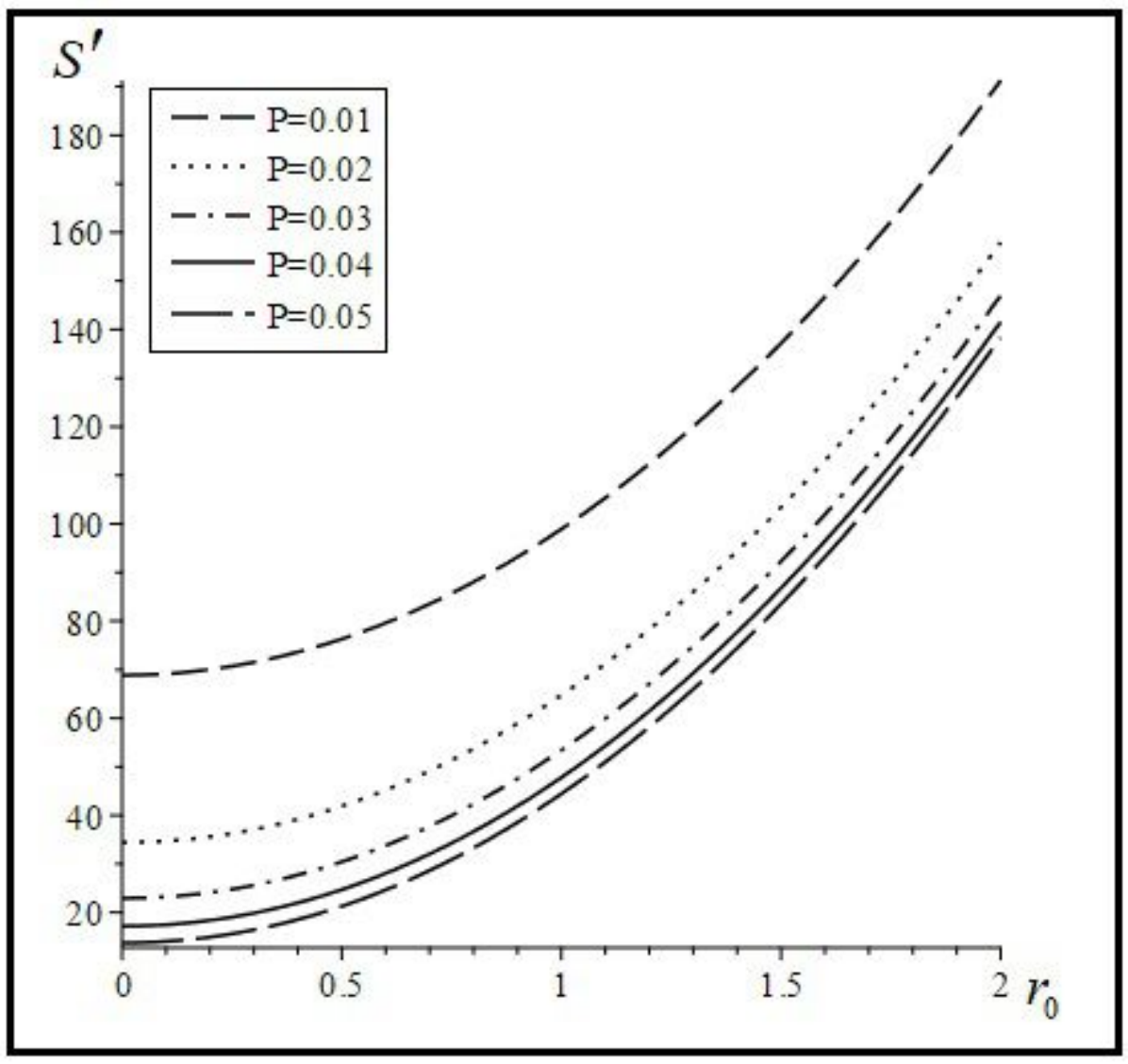}
	\caption{\label{fig:fig1} The corrected entropy of uncharged black hole  vs. $r_{0}$ for $a=0.4$.}
\end{figure}

	In order to investigate some numerical analysis,
	we depict in figure~\ref{fig:fig1} the behavior of uncharged black hole entropy vs. ${r_0}$  and we can see by increasing ${r_0}$  (the position of the event horizon of the black hole) the entropy has increased also by keeping all the other parameters fixed, we have changed the thermodynamic pressure parameter in order to investigate how the black hole entropy is sensitive to parameter $P$  with increasing ${r_0}$. 
	
\begin{figure}[tbp]
	\centering 
	\includegraphics[height=7cm,width=9cm,clip]{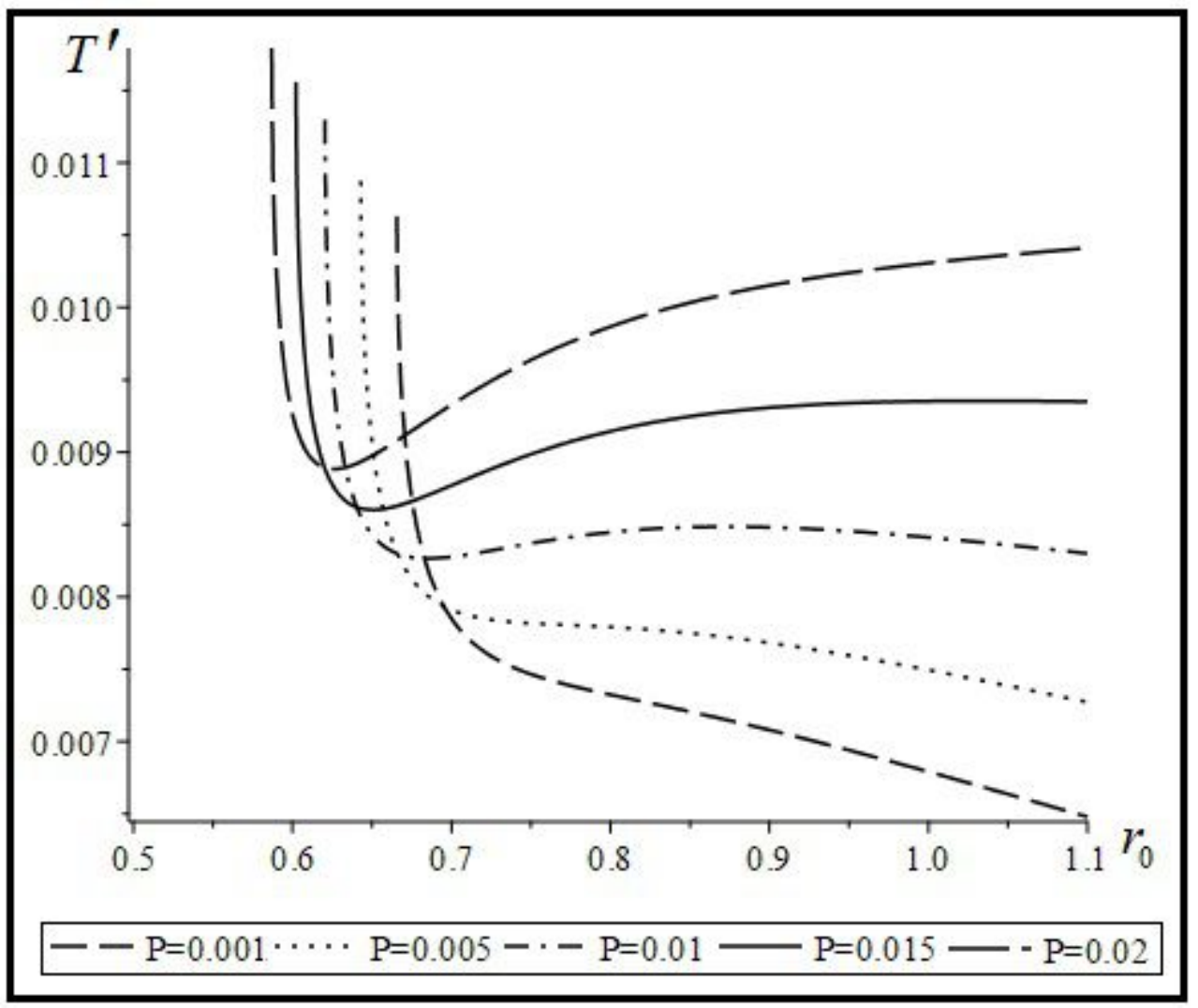}
	\caption{\label{fig:fig2} The corrected temperature of black hole
		vs. ${r_0}$  with $Q=0.5, a= 0.4 $.}
\end{figure}

\begin{figure}[tbp]
	\centering 
	\includegraphics[height=7cm,width=9cm,clip]{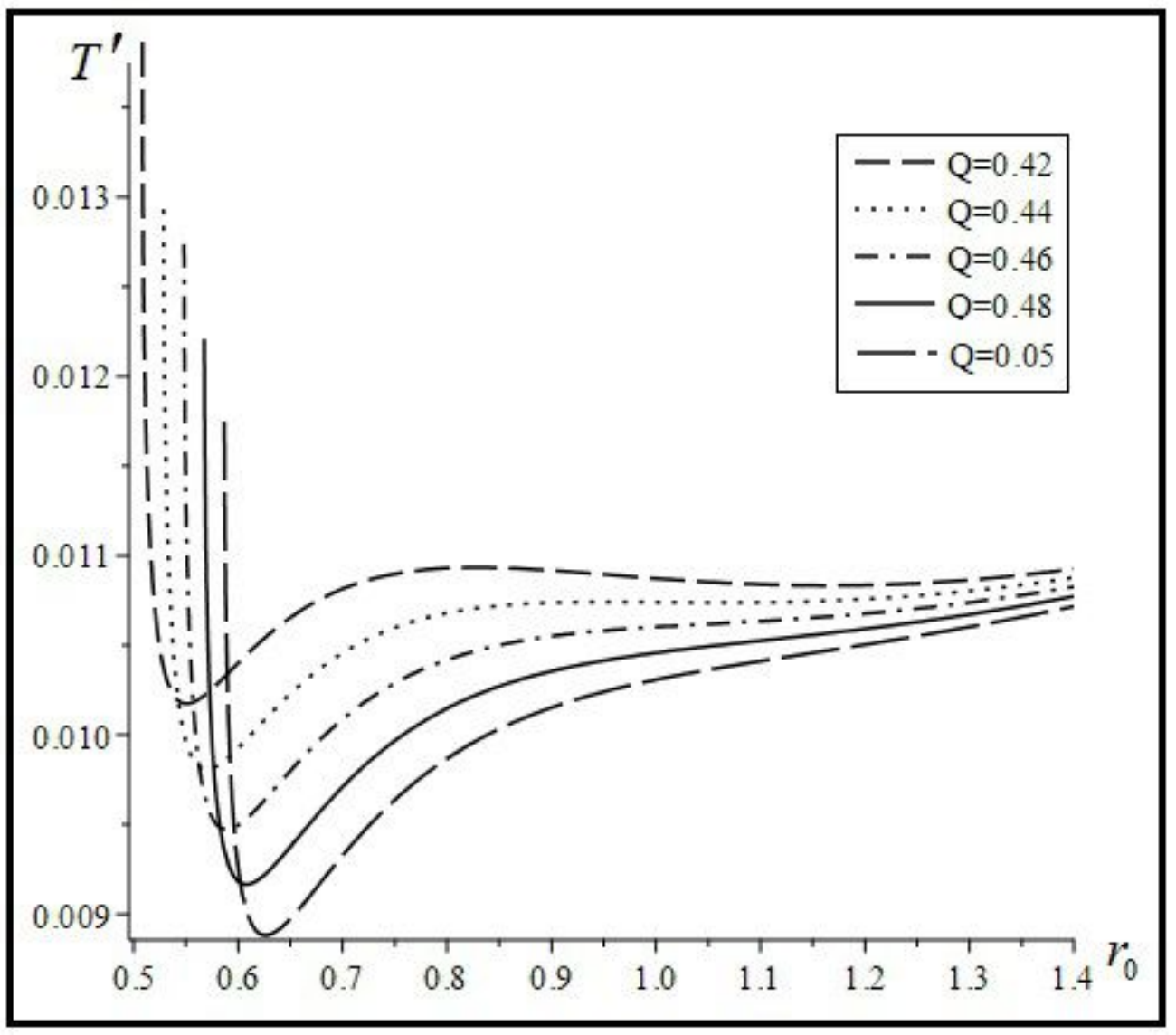}
	\caption{\label{fig:fig3} The corrected temperature of black hole
		vs. ${r_0}$  with $P=0.02, a= 0.4 $.}
\end{figure}

	We have investigated the DSR-GUP corrected temperature of a charged black hole in figures~\ref{fig:fig2} and \ref{fig:fig3}. figure~\ref{fig:fig2} is considered to give a better insight of the corrected black hole temperature vs. the thermodynamic pressure $P$ and it reports the role of the thermodynamic pressure in the temperature spectrum. It is readily observed that the temperature spectrum increases with increasing $P$. Also figure~\ref{fig:fig2} have interesting behavior via  ${r_0}$, as far as there is a sharp decreasing of the temperature of a black hole or a pick in the chart for a certain value of ${r_0}$, by means of the temperature decreases with increasing ${r_0}$ and then it increases.
	We can see from figures~\ref{fig:fig2} and \ref{fig:fig3} that although black hole temperature obtained in view of DSR-GUP increases with increasing  $P$, it decreases with increasing the electric charge $Q$.
	
\begin{figure}[tbp]
	\centering 
	\includegraphics[height=7cm,width=9cm,clip]{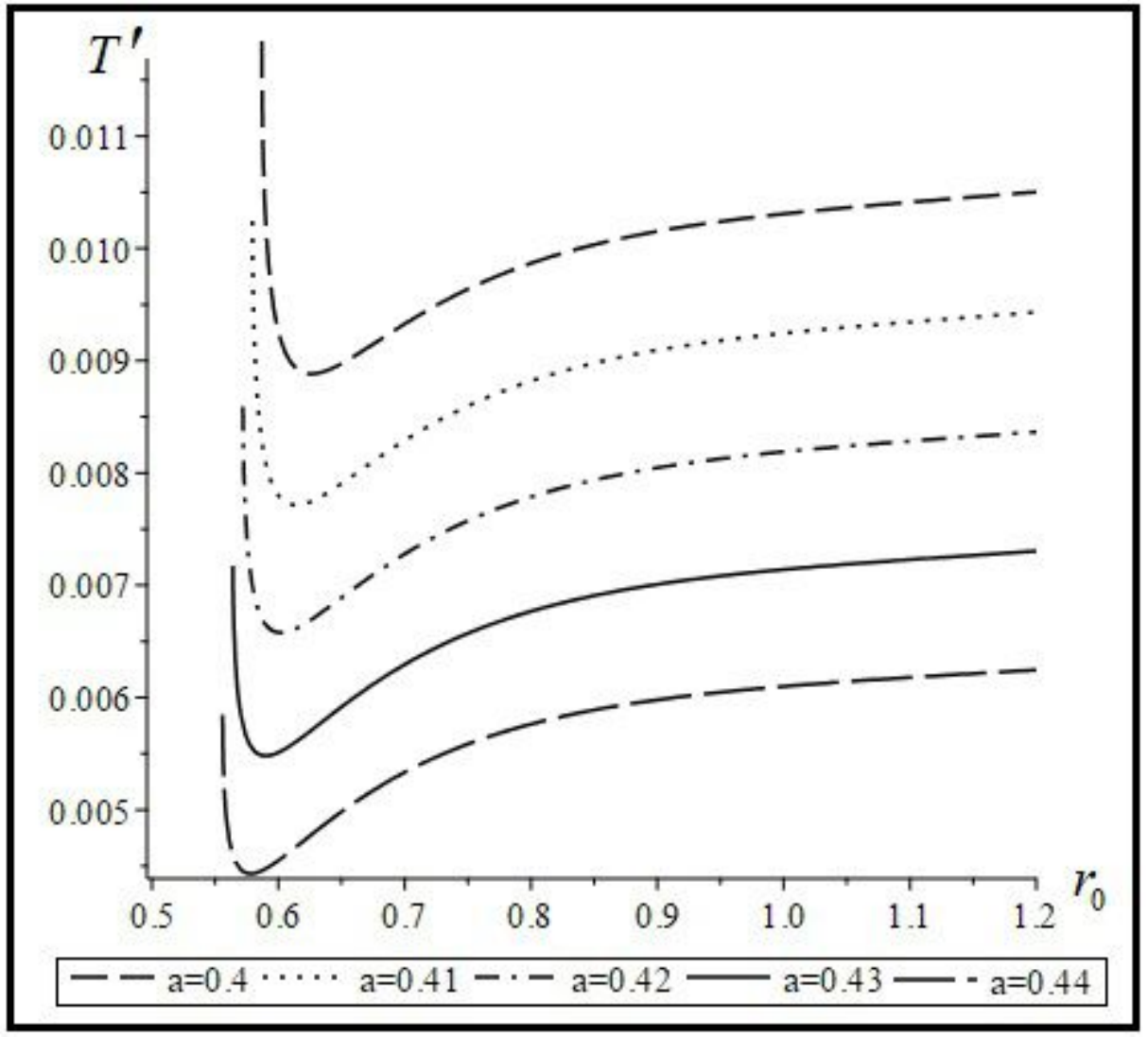}
	\caption{\label{fig:fig4} The corrected temperature of black hole
		vs. ${r_0}$  with $P=0.02, Q= 0.5 $.}
\end{figure}

	By keeping all the other parameters fixed, we have changed a parameter in order to investigate how the black hole temperature are sensitive to parameter $a$  that $0 < a < \frac{1}{2}$  and in figure~\ref{fig:fig4}, one can see the corrected temperature decreases with increasing $a$. 
	
\begin{figure}[tbp]
	\centering 
	\includegraphics[height=7cm,width=9cm,clip]{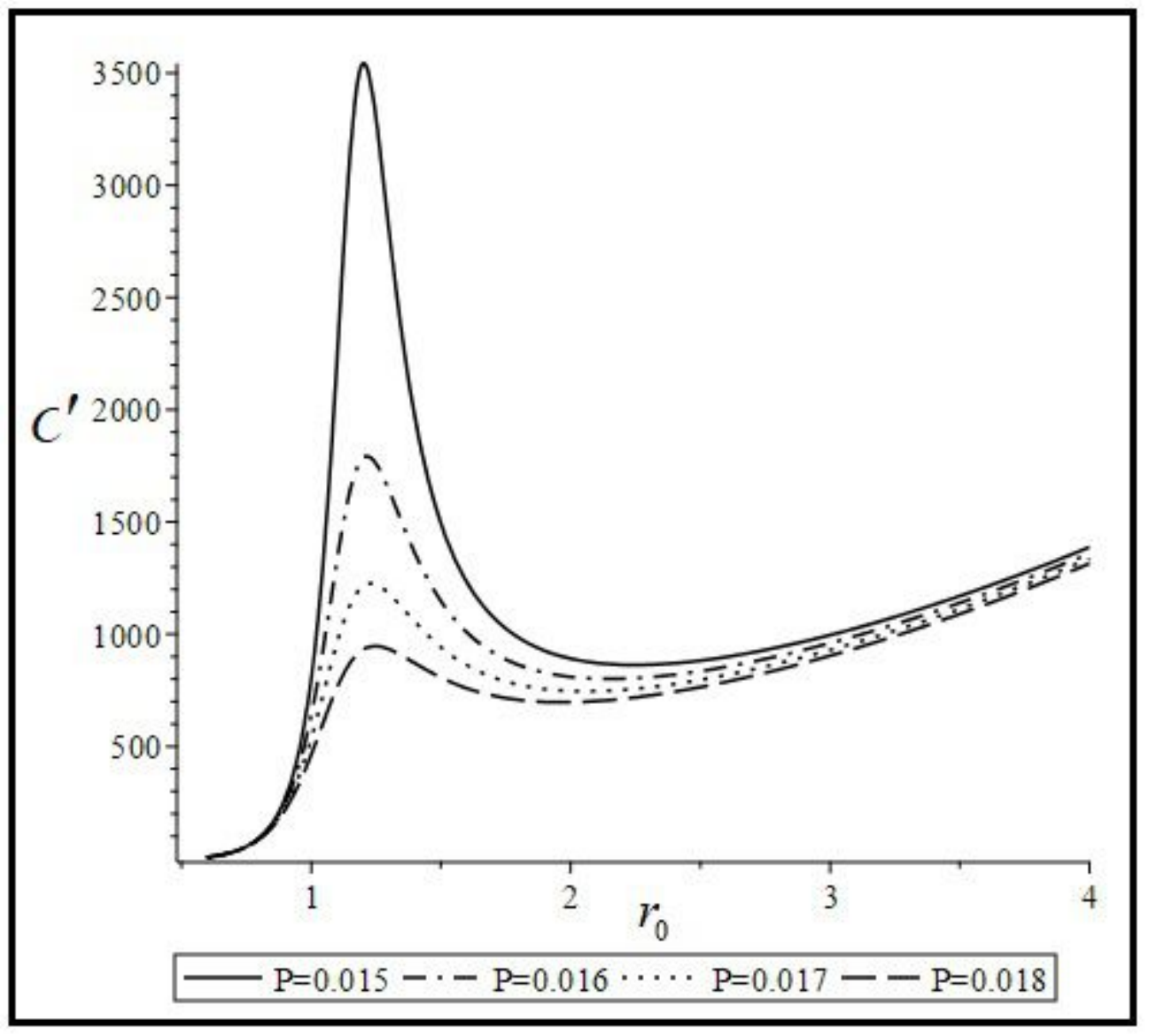}
	\caption{\label{fig:fig5} The heat capacity  vs. ${r_0}$  for $Q=0.5, a=0.4$.}
\end{figure}

	By taking a set of parameter $a=0.4$ and $Q=0.5$, we have investigated the behavior of heat capacity for various values of $P$  vs. $r_{0}$  under the conditions of the DSR-GUP corrected in  figure~\ref{fig:fig5}. 
	
	figure~\ref{fig:fig5} have interesting and weird behavior $C'$  vs. $r_{0}$, as far as there is a sharp increasing of the black hole heat capacity in the chart for a certain value of $r_{0}$, then the heat capacity decreases with increasing $r_{0}$.  Also figure~\ref{fig:fig5} shows that the corrected heat capacity of black hole decreases as $P$  increases. As seen there, DSR-GUP influences on the temperature, the entropy and the heat capacity.

	
	\section{Conclusions}
	\label{sec:Conclusions}
	In this manuscript, by considering the black hole with the special topology $(K=1)$, we have obtained the temperature of the charged AdS black hole under DSR-GUP. One can see obtained results are different of their semiclassical form. Also the corrected entropy has been found for the uncharged black hole and as is shown in figures corresponding to of the corrected forms of entropy, temperature and heat capacity, there is the unique physical critical point.
	
	Our numerical data describe the thermodynamic properties in detail under the position of the event horizon of the black hole  $r_{0}$. As seen in numerical results, DSR-GUP influences on the temperature, the entropy and the heat capacity. The results of our work are comparable with the previous works of other authors in the literature for generalized uncertainty.




\newpage

\end{document}